\documentclass[a4paper,fleqn,usenatbib]{mnras}

\usepackage[T1]{fontenc}
\usepackage{ae,aecompl}

\usepackage{graphicx}	
\usepackage{amsmath}	
\usepackage{amssymb}	

\def\aap{A\&A.}
\def\mnras{MNRAS.}
\def\apjl{ApJL.}
\def\jcap{JCAP.}
\def\prd{PRD}
\def\apj{ApJ.}
\def\apjs{ApJS.}

\def\physrep{Phys. Rep.}

\newcommand{\LCDM}{\rm{\Lambda}CDM}
\newcommand{\Mpc}{\mathrm{~km~s^{-1}~Mpc^{-1}}}

\newcommand{\gp}{\gamma_{\rm{PPN}}}

\title[Determining cosmological-model-independent $H_0$ and post-Newtonian parameter ]{Determining cosmological-model-independent $H_0$ and post-Newtonian parameter with time-delay lenses and supernovae}

\author[Tonghua Liu and Kai Liao ]{
Tonghua Liu$^{1}$ and Kai Liao$^{2}$\thanks{E-mail:liaokai@whu.edu.cn}\\
$^{1}$School of Physics and Optoelectronic Engineering, Yangtze University, Jingzhou, 434023, China\\
$^{2}$Department of Astronomy, School of Physics and Technology, Wuhan University, Wuhan 430072, China\\
}

\date{Accepted XXX. Received YYY; in original form ZZZ}

\pubyear{2023}
\begin{document}
\maketitle

\begin{abstract}
Strong gravitational lensing provides a natural opportunity to test General Relativity (GR). We propose a model-independent method for simultaneous constraining on Hubble constant ($H_0$) and  post-Newtonian parameter ($\gp$) using strong lensing systems and observational SNe Ia.  The time-delay measurements from strong lesning can directly determine the Hubble constant, and the lens distance inferred from the spectroscopic measurement of the stellar kinematics of the deflector galaxy can help us to constrain the post-Newtonian parameter.  We seek the Pantheon dataset and reconstruct unanchored distances using Gaussian process regression to achieve the cosmological model-independent GR testing instead of assuming a specific model, which can reduce possible bias on GR testing and  measurement of Hubble constant. Combining the reconstructed unanchored distances and the four H0LiCOW lenses datasets, our results are $H_0=72.9^{+2.0}_{-2.3} \Mpc$ and $\gp=0.89^{+0.17}_{-0.15}$.  All the lenses show that there is no obvious evidence to support GR deviation within observational uncertainties. In the subsequent analysis, we consider a ratio of distance ${D_{\Delta t}}/{D^{'}_{d}}$ method to further avoid the influence of $H_0$ on GR testing. The results show that, except J1206 within the $\sim1.2\sigma$ observational uncertainty, the remaining 3 lenses support GR holds within the $1\sigma$ observational uncertainties.

\end{abstract}

\begin{keywords}
 cosmology: cosmological parameters - distance scale - gravitational lensing: strong
\end{keywords}

\section{Introduction}
The modern theory of cosmology is based on two pillars, Einstein's theory of General Relativity (GR) and the cosmological principle. The former was the first to equate the gravitational field with the curvature of space-time and was extremely successful in describing the gravitational interaction between matter. The latter describes that our Universe is homogeneous and isotropic on large scales.
The Lambda cold dark matter ($\LCDM$) model is consistent with most popular observational evidences, such as the observations of type Ia supernovae (SNe Ia) \citep{2007ApJ...659...98R}, Cosmic Microwave Background Radiation (CMB) \citep{2016A&A...594A..13P}, and regarded as the standard cosmological model.
However, there is a recent tension of 4.4$\sigma$ or more between the Hubble constant inferred by CMB observations under the assumption of a flat $\LCDM$ model \citep{2020A&A...641A...6P} and its value measured through Cepheid-calibrated distance ladder by the \textit{Supernova $H_0$ for the Equation of State} collaboration (SH0ES) \citep{2019ApJ...876...85R}. Such inconsistency could be caused by unknown systematic errors in astrophysical observations or reveal new physics significantly different from $\Lambda$CDM model.
In a recent work, \citet{2022ApJ...938..111B} reanalyzed and re-calibrated the photometric systems in the Pantheon+sample of SN Ia, including those in the SH0ES distance-ladder measurement of $H_0$, and suggested that supernova calibration is not currently capable of resolving Hubble tension.

As an important prediction of GR, gravitational lensing is a powerful tool to study the velocity dispersion function of early-type galaxies \citep{2008MNRAS.384..843M,2021MNRAS.503.1319G,2007ApJ...658L..71C}, the distribution of dark matter \citep{2022A&A...659L...5C,2011ApJ...727...96R,1993ApJ...407...33M,2009ApJ...706.1078N}, and cosmological parameters \citep{2014ApJ...788L..35S,2017MNRAS.465.4914B,2019ApJ...886...94L,2019MNRAS.488.3745C}.  More importantly, the time-delay measurements between multiple images from strong gravitational lensin  provide a valuable opportunity for determination of $H_0$ \citep{1964MNRAS.128..307R}. In the milestone work of \citet{2020MNRAS.498.1420W}, the six gravitationally lensed quasars with well-measured time delays were jointly used to constrain the $H_0$ by the $H_0$ Lenses in COSMOGRAIL's Wellspring (H0LiCOW) collaboration. Assuming a flat $\Lambda$CDM model, the H0LiCOW collaboration reported the results on $H_0$ with these six lensed quasars,  $H_0=73.3^{+1.7}_{-1.8}$ $\Mpc$, which is consistent with local measurements from SN Ia, but in $3.1\sigma$ tension with CMB observations. However, it needs to be stressed that both the Planck and H0LiCOW inferred $H_0$ values are based on GR plus $\LCDM$ model, that inspires us to investigate the validity of GR with cosmological model-independent way.  The validity of GR can be verified by constraining the parametrized post-Newtonian (PPN) parameter $\gp$ (since GR predicts exactly $\gp\equiv1$), which describes the spatial curvature generated by an unit rest mass. In recent years, especially on the solar system scale, many tests on GR have made great achievements in extremely high precision (see review \citep{2014LRR....17....4W} for more works about testing GR).  However, testing GR at extra-galactic scale is still not very precise. For instance, there is only $\sim 20\%$ precision on $\gp$ at $10-100$ Mpc scales by using the joint measurements of weak gravitational lensing and redshift-space distortions \citep{2013MNRAS.429.2249S,2016MNRAS.456.2806B}. At cosmological scales, strong gravitational lensing systems provide an effective way to probe deviation of GR \citep{2006PhRvD..74f1501B,2017ApJ...835...92C,2022ApJ...927...28L,2022ApJ...927L...1W}.
Recently, the work by \citet{2018Sci...360.1342C} used a nearby strong gravitational lens ESO 325-G004 to test GR and reported the constrained results on $\gp=0.97\pm0.09$ at $1\sigma$ confidence level (C.L).
In further research, \citet{2020MNRAS.497L..56Y} proposed a new methodology through time-delay measurements combined with the stellar kinematics of the deflector lens from strong lensing to simultaneously constrain $H_0$ and $\gp$, and showed that there is no obvious deviation from the GR with the result $\gp=0.87^{+0.19}_{-0.17}$.  However, it should be emphasized here that these work are cosmological model-dependent (in $\LCDM$ model). The testing GR should be done without invoking any particular background cosmology model in order to reduce potential bias from the forms of the parametric or model assumptions.

Inspired by above, we propose a cosmological model-independent method to constrain the Hubble constant and PPN parameter simultaneously at cosmological scales using four strongly lensed quasars published by H0LiCOW with both time delay distance and lens distance. For the distance information required to constrain the PPN parameter, we seek for unanchored (or relative) distances from SN Ia observations using a Gaussian process (GP) regression.
Intuitively,  the Time-Delay Strong Lensing (TDSL) measurements can directly determine the Hubble constant, and the lens distance inferred from the spectroscopic measurement of the stellar kinematics of the deflector galaxy can help us to constrain the post-Newtonian parameter. This paper is organized as follows: In Section \ref{sec2} we introduce the methodology and H0LiCOW lensing data including reconstructed $H_0D^L$ using GP regression. The constrained results on $H_0$ and $\gp$ and discussions are given in Section \ref{sec3}. We conclusion our result in Section \ref{sec4}. The natural units of $c=G=1$ are adopted throughout this work.

\section{Methodology and H0LiCOW lensing data}\label{sec2}
\subsection{Distances inferred from H0LiCOW program}

In the limit of a weak gravitational field, the Schwarzschild line element of space-time can be expressed as
\begin{equation}
ds^2=-\big(1+2\Psi\big) d t^2+\big(1-2\Phi\big)d  r^2+r^2d\Omega^2,
\label{eq:ds}
\end{equation}
where $\Psi$ is the Newtonian potential and $\Phi$ represents the spatial curvature potential, and $\Omega$ is the angle in the invariant orbital plane.  The ratio $\gp=\Phi/\Psi$ denotes as the PPN parameter, which describes the spatial curvature generated per unit mass. It should be emphasized  that  the PPN parameter $\gp$ is predicted to be one or $\Psi=\Phi$. In this work, we assume that $\gp$ is a constant at the lens galaxy scales.

The crucial idea of using strong lensing systems to test gravity is through two different mass measurements, i.e., gravitational mass inferred from the lensing image, and dynamical mass obtained from the spectroscopic measurement of stellar kinematics of the deflector galaxy. The motion of non-relativistic matters (usually made up of baryonic matter and dark matter) is governed by the Newtonian gravitational potential $\Psi$, which obeys Poisson equation. However, the motion of relativistic particles is sensitive to both potentials.  Testing $\gp$ requires observing the motion of relativistic and non-relativistic particles around the same massive object. Thus, strong lensing systems provide a natural laboratory to test gravity and further measure the PPN parameter $\gp$.
The difference between dynamical mass and lensing mass can be directly used for difference between $\Psi$ and $\Psi_{+}=\frac{\Psi+\Phi}{2}=(\frac{1+\gp}{2})\Psi$ (namely Weyl potential).  In the framework of PPN, the deflection angle contains the lensing mass information, $\alpha_{\rm PPN}(\theta)=(\frac{1+\gp}{2})\alpha_{\rm{GR}}(\theta)$, and effective
lensing potential (the integral of the Weyl potential along the line-of-sight) rescales as $\psi_+=(\frac{1+\gamma_{\rm{PPN}}}{2})\psi$, as well as convergence field $\kappa'=(\frac{1+\gamma_{\rm{PPN}}}{2})\kappa$. It is worth noting that the deflector angle is directly related to the cosmological distance, so the PPN parameter and the cosmological distance are highly degenerated. This is one of the main limitations of using strong lensing systems to test gravity.
To break this degeneracy, additional data needs to be taken into account, either the cosmological or the gravitational one. The time delay measurements are able to break the degeneracy alone.

For a given strong lensing system, quasar acting as background source, time delays between multiple images can be measured from variable AGN light curves, and determined by both the geometry of the Universe as well as the gravitational potential of the lensing galaxy through \citep{1964PhRvL..13..789S}
\begin{equation}
\Delta t=D_{\Delta t}\left[\phi(\theta_{\rm A},\beta)-\phi(\theta_{\rm B},\beta)\right]=D_{\Delta t}\Delta\phi_{\rm AB}(\xi_{\rm lens}),
\end{equation}
where $\phi(\theta,\beta)=\left[{(\theta-\beta)^2}/{2}-\psi(\theta)\right]$ is the Fermat potential at images, $\beta$ is the source position, $\xi_{\rm lens}$ is the lens model parameter. The cosmological background is reflected in so-called "time delay distance" $D_{\Delta t}=(1+z_{\rm d}){D_{\rm d}D_{\rm s}}/{D_{\rm ds}}$, which is inverse proportional to $H_0$. The key here is the Fermat potential difference $\Delta\phi_{\rm AB}(\xi_{\rm lens})$, which can be reconstructed by high-resolution lensing imaging from space telescopes. As we mentioned above, in the PPN framework,  the inferred mass parameters are rescaled by a factor of $(1+\gp)/2$. Therefore, we denote the actually inferred lens model parameters in the Fermat potential as $\xi^{'}_{\rm lens}$. We rewrite the time-delay distance as
\begin{equation}
D_{\Delta t}=(1+z_{\rm d})\frac{D_{\rm d}D_{\rm s}}{D_{\rm ds}}=\frac{\Delta t_{\rm AB}}{\Delta\phi_{\rm AB}(\xi'_{\rm lens})}\,.
\label{eq:ddt}
\end{equation}
This is the first distance we need. It can be obtained from both the measurements of time delay and the Fermat potential reconstructed with parameter $\xi^{'}_{\rm lens}$.

On the other hand, the stellar kinematics of lensing galaxies are only sensitive to the Newtonian potential $\Psi$, which are independent of PPN parameters. It can also be obtained from the modeling of the kinematic observable in lensing galaxies. The modeling of the stellar kinematic in lensing galaxies by the H0LiCOW collaboration is quite mature, and here we give a brief introduction, thus provide the reader with a detailed background. The dynamics of stars with the luminosity density distribution of lenses $\rho_*(r)$
in the gravitational potential $\Psi$ follows the Jeans equation. In the limit of a relaxed (vanishing time
derivatives) and spherically symmetric system, with the only distinction between radial ($\sigma_r$) and tangential ($\sigma_t$) dispersions, the anisotropic Jeans equation is
\begin{equation}
\frac{\partial(\rho_*\sigma_r^2)}{\partial r}+\frac{2\beta_{\rm{ani}}(r)\rho_*\sigma_r^2}{r}=-\rho_*\frac{\partial\Psi}{\partial r}\,,
\label{eq:jeans}
\end{equation}
where $\beta_{\rm{ani}}(r)\equiv 1-\frac{\sigma_t^2}{\sigma_r^2}$ is the stellar distribution anisotropy. The luminosity-weighted projected velocity dispersion $\sigma_s$ is $I(R)\sigma_s^2=2\int_R^\infty \left(1-\beta_{\rm{ani}}(r)\frac{R^2}{r^2}\right)\frac{\rho_*\sigma_r^2rdr}{\sqrt{r^2-R^2}}$ \citep{2010ApJ...711..201S},
where $I(R)$ is the projected light distribution and $R$ is the projected radius. Considering observational conditions, the luminosity-weighted line-of-sight velocity dispersion $\sigma_v$ within an aperture $\mathcal{A}$ is real observations, which is given by $\sigma_v^2=\frac{\int_\mathcal{A}[I(R)\sigma_s^2*\mathcal{P}]dA}{\int_\mathcal{A}[I(R)*\mathcal{P}]dA}$, where $\mathcal{P}$ is point spread function (PSF) convolution of the seeing.
The prediction of the stellar kinematics requires a three-dimensional stellar density $\rho_*$ and mass $M(r)$ profile. In terms of imaging data, the information about the parameters
of the lens mass surface density with parameters mass $\xi_{\rm{lens}}$ and the
surface brightness of the lens with parameters $\xi_{\rm{light}}$ can be extract.
Finally, the prediction of any $\sigma_v$ from any model can be decomposed into cosmological part $D_s/D_{ds}$ and stellar kinematics part $J(\xi_{\rm{lens}},\xi_{\rm{light}},\beta_{\rm{ani}})$~\citep{2019MNRAS.484.4726B}.
The function $J$ captures all of the model components calculated from the sky angle (from the imaging data) and the anisotropy distribution of the stellar orbit (from the spectroscopy).
This allows us to obtain the distance ratio $D_s/D_{ds}$ from the well-measured velocity dispersion, independent of the cosmological model and time delays, but still relies on the lens model $\xi_{\rm{lens}}$ (not the $\xi^{'}_{\rm{lens}}$ under PPN) \citep{2016JCAP...08..020B,2019MNRAS.484.4726B}
\begin{equation}
\frac{D_{\rm s}}{D_{\rm ds}}=\frac{\sigma_v^2}{c^2J(\xi_{\rm{lens}},\xi_{\rm{light}},\beta_{\rm{ani}})}\;.
\label{eq:gr_dsdds}
\end{equation}
The lens model parameter in $J$ is the ``unrescaled'' $\xi_{\rm{lens}}$. Here, we use $\xi'_{\rm{lens}}$ to replace $\xi_{\rm{lens}}$, the corresponding distance ratio shall also be rescaled, as
\begin{equation}
\frac{2}{1+\gamma_{\rm{PPN}}}\frac{D_{\rm s}}{D_{\rm ds}}=\frac{\sigma_v^2}{c^2J(\xi'_{\rm{lens}},\xi_{\rm{light}},\beta_{\rm{ani}})}\,.
\label{eq:dsdds1}
\end{equation}
Furthermore, we can define rescaled deflector galaxy distance $D_{\rm d}'=\frac{1+\gamma_{\rm{PPN}}}{2}D_{\rm d}$.
By combining Eqs.~(\ref{eq:ddt}) and (\ref{eq:dsdds1}), we obtain \citep{2016JCAP...08..020B,2019MNRAS.484.4726B}
\begin{eqnarray}
D_{\rm d}'=\frac{1}{1+z_{\rm d}}\frac{c\Delta t_{\rm AB}}{\Delta \phi_{\rm AB}(\xi'_{\rm{lens}})}\frac{c^2J(\xi'_{\rm{lens}},\xi_{\rm{light}},\beta_{\rm{ani}})}{\sigma_v^2}\,.
\label{eq:ddp}
\end{eqnarray}
This is the second distance we need. More details for these two distance obtained from strong lensing systems refer to work \citep{2020MNRAS.498.1420W} and references therein.

Thanks to the H0LiCOW collaboration, four lenses (namely RXJ1131-1231 \citep{Suyu13,2014ApJ...788L..35S}, PG1115+080 \citep{Chen19}, B1608+656 \footnote{This len was given in the form of skewed log-normal distribution, due to  the absence
of blindly analysis with respect to the cosmological parameters} \citep{2010ApJ...711..201S,Jee19},
J1206+4332 \citep{2019MNRAS.484.4726B}) posterior distributions with both angular diameter distances of lens $D_d$ and time delay distances $D_{\Delta t}$ have been published, making it easier to use them. There are available on the website of H0LiCOW \footnote{http://www.h0licow.org}.
The redshifts of both lens and source, the time delay distances, and the angular diameter distance to the lenses for these lensed quasar systems are summarized in Table 2 of \citet{2020MNRAS.498.1420W}.
For more relevant work by using these lensing systems, we refer the reader to see the literature \citep{2021MNRAS.503.1096D,2022A&A...668A..51L,2022ApJ...927..191B,2021A&A...656A.153S,2015ApJ...800...11L,2015A&A...580A..38R,2020ApJ...895L..29L,2022ApJ...939...37L}.

The next problem is the acquisition of these two distance information.
However, since $D_d$ depend on the cosmological model, $D_{\Delta t}$ also change in different cosmological models. The traditional method is to assume a particular cosmological model, such as a flat $\LCDM$ model, but it is cosmological model-dependent.
In this work,  we follow the reconstructed method used in work \citep{2019ApJ...886L..23L}, and seek for current SN Ia observations to determine distances, even though the observations of SN Ia are anchored relative distances.

\subsection{Unanchored distance from observations of SNe Ia using GP regression}

As the most explosive variable source in Universe, due to the nature of SNe Ia (as standard candles), they were regarded as powerful cosmological probes. It was the observations of SNe Ia that led to the discovery of the accelerating expansion of Universe. We use recent Pantheon dataset, consisting of 1048 SNe Ia spanning the redshift range $0.01<z<2.3$ \citep{2018ApJ...859..101S}.  To combine the Pantheon SNe and the H0LiCOW strong lenses datasets, we generate samples of the unanchored luminosity distance $H_0D^L$ from the posterior of the Pantheon dataset. In order to perform the posterior sampling in a cosmological model-independent way, the GP regression~\citep{Holsclaw,Holsclaw1,Keeley0,ShafKimLind,ShafKimLind2,
2019JCAP...12..035K} is considered here by using the code \texttt{GPHist}\footnote{https://github.com/dkirkby/gphist} \citep{GPHist}. GP regression is powerful tool for reconstruction of function since the regression occurs in an infinite-dimensional function space without overfitting problem \citep{Keeley0}. GP regression works by generating a large sample of functions $\gamma(z)$ determined by the covariance function.
The covariance between these functions can be described by a kernel function. We adopt a squared-exponential kernel to parameterize the covariance
\begin{equation}
    \langle \gamma(z_1)\gamma(z_2) \rangle = \sigma_f^2 \, \exp\{-[s(z_1)-s(z_2)]^2/(2\ell^2)\},
\end{equation}
with hyperparameters $\sigma_f$ and $\ell$ that are marginalized over.  The $\gamma(z)$ is a random function inferred from the distribution defined by the covariance, and we adopt $\gamma(z) = \ln([H^{\rm fid}(z)/H_0]/[H(z))/H_0])$ to generate expansion histories $H(z)/H_0$ by using Pantheon dataset.
Here $H^{\rm fid}(z)/H_0$ is chosen to be the best fit $\Lambda$CDM model for the Pantheon data and serves the role of the mean function for GP regression.  Since the final reconstruction result is not completely independent of the mean function, it has some influence on the final reconstruction result because the value of the hyperparameter helps to track the deviation from the mean function. And the true model should be very
close to flat $\LCDM$, hence, our choice for the mean function is very reasonable \citep{ShafKimLind,ShafKimLind2,Aghamousa2017}.

\begin{figure}
 \includegraphics[width=\columnwidth,angle=0]{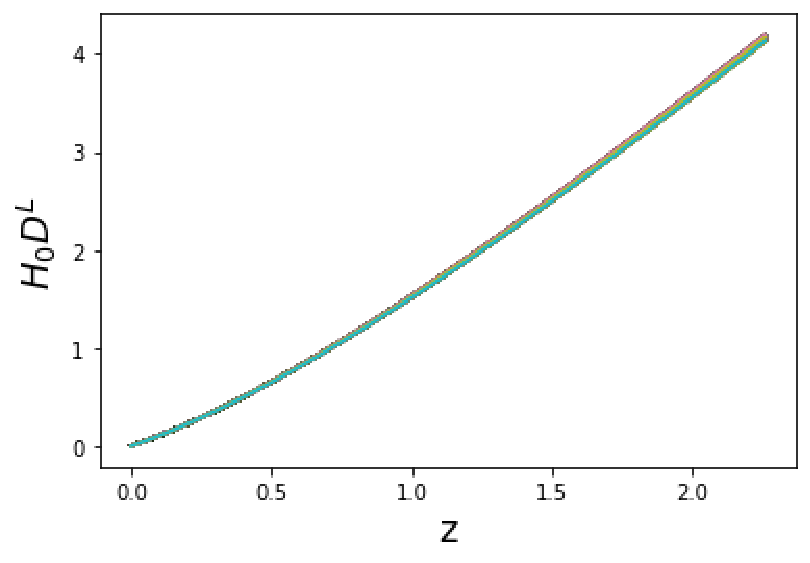}
  \caption{
The unanchored luminosity distance curves $H_0D^L(z)$ reconstructed from the Pantheon SN Ia data for a representative sample of the 1000 GP realizations.
  } \label{fig:sn}
\end{figure}

To highlight the purpose of our work, i.e., the Hubble constant and $\gp$, we assume a flat universe here.
With the reconstructed expansion history $H(z)/H_0$, the unanchored SN luminosity distances can be calculated
\begin{equation}
    H_0 D^L(z) = (1+z) \int^z_0 dz'/[H(z')/H_0]\ .
\end{equation}
The 1000 unanchored luminosity distance curves $H_0D^L(z)$ reconstructed from the SN data are shown in Fig.~\ref{fig:sn}. It shows the shape of the cosmic distance-redshift relation of Pantheon data very well.
It should be noted that the redshift of SN dataset well covers range of four strong lensing system redshifts, so the redshift range needn't extrapolation.

\subsection{Simultaneous measurements on Hubble constant and PPN parameters}
\begin{figure*}
 \centering
\includegraphics[width=7.8cm,height=6.4cm]{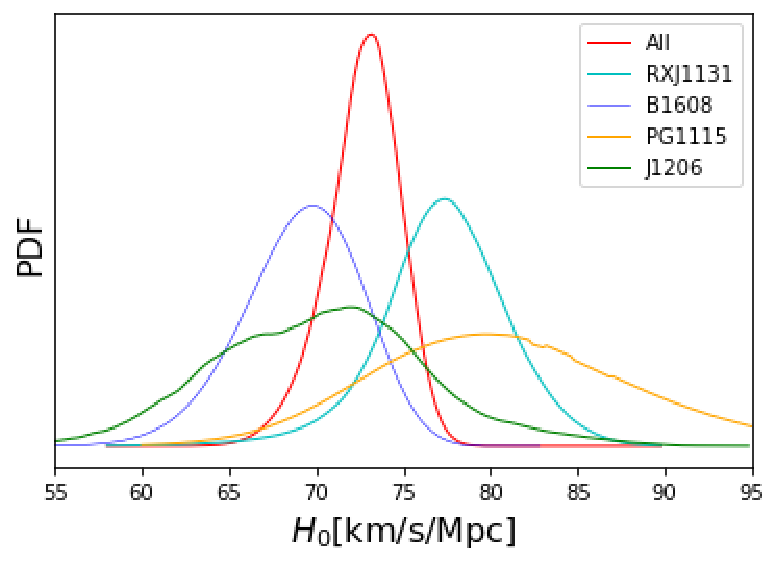}
\includegraphics[width=7.8cm,height=6.4cm]{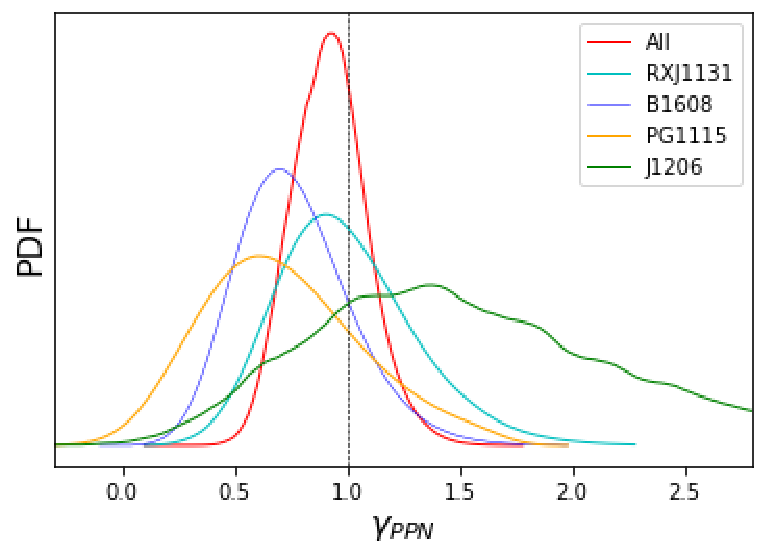}
\caption{The simultaneous constraints of $H_0$ (left panel) and $\gamma_{\rm{PPN}}$ (right panel) from four of the H0LiCOW lenses. The dashed line is $\gamma_{\rm{PPN}}=1$ predicted by GR. }\label{fig:result_h0licow}
\end{figure*}

In summary, combining the three measurements, namely optical lensing image, deflector spectroscopies as well as time delays, two distances ($D_{\Delta t}, D_{\rm d}'$) can be inferred, simultaneously, based on Eqs.~(\ref{eq:ddt}) and (\ref{eq:ddp}).
It should be stressed that there is only deflector galaxy distance ($D_{\rm d}'$) carried the information of the PPN parameter, while time delay distance $(D_{\Delta t})$ is sensitive to $H_0$.
Thus, the combination of $D_{\Delta t}$ and $D_{\rm d}'$ directly provides a new way for simultaneous measurement of $H_0$ and $\gp$.

The steps for simultaneous constraining $H_0$ and $\gp$ are summarized as follows:
\begin{enumerate}
\item
    Draw 1000 unanchored luminosity distances curves $H_0D^L$ from SN data, and convert to unanchored angular diameter distances $H_0D^A$ for serving the strong lensing systems.
\item
    Calculate 1000 values of $H_0D_d$ at the lens from the 1000 $H_0D^A$ curves for serving four posterior distributions of lens from H0LiCOW program; Adopt same procedure for source redshifts of the four strong lens systems to calculate 1000 values of $H_0D_s$; Combine these $H_0D_d$ and $H_0D_s$ to calculate $H_0D_{\Delta t}$ for each system using $H_0D_{\Delta t}=(1+z_d)(H_0D_d)(H_0D_s)/(H_0D_{ds})$\footnote{We use the standard distance relation to obtain the angular diameter distance between the lens and the source \citep{Weinberg1972} for a spatially flat universe $D_{ds}=D_s-[(1+z_d)/(1+z_s)]D_d$.};
\item
    Compute the likelihood, for each of the 1000 realizations, from the H0LiCOW's $D_{\Delta t}$ and $D_{d}$ data for each lens system for many values of $H_0$ and $\gp$;
\item
    Multiply the four likelihoods to form the full likelihood for each realization, for each values of
    $H_0$ and $\gp$;
\item
     Marginalize over the realizations to form the posterior distributions of $H_0$ and $\gp$.
\end{enumerate}

\section{Results and discussions}\label{sec3}
Working on four posteriors of $D_{\Delta t}$ and $D_{d}$ published by H0LiCOW program and combining reconstructed unanchored SN dataset, we obtain the final distributions for Hubble constant $H_0$ and PPN parameter $\gp$.
Our model-independent constraints are $H_0=72.9^{+2.0}_{-2.3} \Mpc$ and $\gp=0.89^{+0.17}_{-0.15}$ (median value plus the $16^{th}$ and $84^{th}$ percentiles around this) for combination of all lenses.
The one-dimensional posterior distributions are shown in Fig.~\ref{fig:result_h0licow}.
The numerical constraint results for four individual lenses can be found in Table~\ref{h0andgp}.
One can see that constraint results of all the lenses show that GR is supported within observational uncertainties, and there is no obvious evidence of GR deviation. This can be compared to the results of \citet{2020MNRAS.497L..56Y} of $H_0=73.65^{1.95}_{-2.26} \Mpc$ and $\gp=0.87^{+0.19}_{-0.17}$ within a
flat $\LCDM$ for combining all lenses, our results are agreement with their results, which support the robustness of our method.  Our results on $H_0$ are consistent with the time delay and time delay plus SN results \citep{2019PhRvL.123w1101C,2020MNRAS.498.1420W,2019A&A...628L...7T} made within specific cosmological models or with polynomial fitting of the distance relation.
However, it should be stressed that, comparing with assuming a specific model,  combinations of strong lensing and current astronomical probe such as SN Ia can reduce possible bias in our work. More importantly, there is no significant increase in uncertainty.

\begin{table}
\renewcommand\arraystretch{1.3}
\caption{ Summary of the constraints on the Hubble constant $H_0$ and $\gp$ from four of the H0LiCOW lenses.}

\begin{center}
\begin{tabular}{l| c| c }
\hline
\hline
Lenses  & $H_0 (\Mpc)$ &$\gp$  
\\
\hline
All  & $72.9^{+2.0}_{-2.3}$ & $0.89^{+0.17}_{-0.15}$ \\
\hline
RXJ1131  & $77.6^{+3.3}_{-3.4}$ & $0.94^{+0.33}_{-0.27}$ \\
\hline
B1608  & $69.5^{+3.2}_{-3.3}$ & $0.73^{+0.27}_{-0.23}$ \\
\hline
PG1115  & $80.4^{+7.8}_{-7.1}$ & $0.67^{+0.41}_{-0.33}$ \\
\hline
J1206  & $71.8^{+7.2}_{-6.8}$ & $1.44^{+0.73}_{-0.61}$ \\
\hline
\hline
\end{tabular}
\end{center}\label{h0andgp}
\end{table}

\begin{figure}
\includegraphics[width=7.8cm,height=6.4cm]{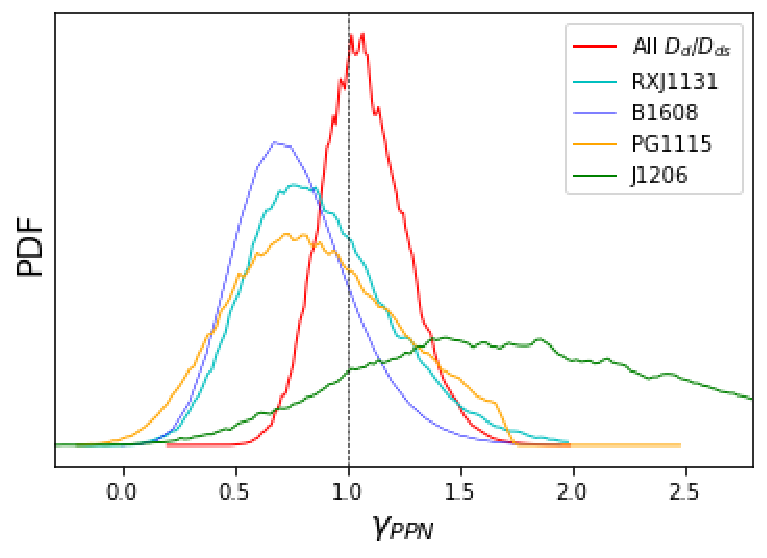}
\caption{The one-dimensional posterior distributions of $\gamma_{\rm{PPN}}$ using distance ratio method with four H0LiCOW lenses. The dashed line is $\gamma_{\rm{PPN}}=1$ predicted by GR.}
\end{figure}\label{radiofig}

\begin{table}
\renewcommand\arraystretch{1.3}
\caption{Summary of the constraints on $\gp$ using the distance ratio method from four H0LiCOW lenses.}

\begin{center}
\setlength{\tabcolsep}{1.5mm}{
\begin{tabular}{l| c|c|c|c|c}
\hline
\hline
Lenses  & All  & RXJ1131& B1608 &PG1115  &J1206 
\\
\hline
$\gp$ & $1.07^{+0.20}_{-0.17}$  & $0.85^{+0.34}_{-0.28}$ & $0.73^{+0.27}_{-0.23}$ & $0.84^{+0.30}_{-0.36}$ & $1.78^{+0.87}_{-0.67}$\\
\hline
\hline
\end{tabular}}
\end{center}\label{radio}
\end{table}

On the other hand, Hubble constant and PPN parameter are in fact completely degenerate together, this can also be seen in work of \citet{2020MNRAS.497L..56Y}. From a theoretical point of view, PPN parameter is encoded in the inferred angular diameter distance to the lens, $D'_{\rm d}=D_{\rm d}(1+\gp)/2$, which leads that PPN parameter and cosmological distance (related to Hubble constant) are degenerated. For instance, if $\gp$ increase, it means the enhancement of gravitational force. One also can keep gravity unmodified, but change the corresponding distances. To try to break this degeneracy and do not assume any value for $H_0$, we can also consider ratios of distances ${D_{\Delta t}}/{D^{'}_{d}}=\frac{2}{1+\gp}{D_{\rm s}}/{D_{\rm ds}}$,
which are independent of $H_0$. The final results are displayed in Fig~\ref{radiofig}. We see that the mean values of $\gp$ for four individual lenses have a little changes (though not significant), the numerical results are shown in Table~\ref{radio}. For combination of all lenses, the $\gp$ parameter is even closer to one. However, for the lens J1206, the corresponding constraint result is $\gp=1.78^{+0.87}_{-0.67}$.  Although GR is somewhat deviated within the $1\sigma$ confidence level, GR is still supported  within the $\sim1.2\sigma$ confidence level. As pointed out in work of \citet{2020A&A...639A.101M}, the dispersion measurements do not play a significant role in the $H_0$ estimation of the H0LiCOW analysis, except for the lens J1206. In other word, in the analysis of the distance ratios, since we have not given any value for $H_0$, the degeneracy between $H_0$ and $\gp$ causes a shift in $\gp$. In addition, for the lens B1608, the constraint of $\gp$ using the distance ratio does not change at all, the $D_{\Delta t}$ and $D_{d}$ of this lens are completely independent, because of the lack of blind analysis of this lens. Since no new data has been added, we do not expect any improvement in precision, and our results show this point.

\section{Conclusion}\label{sec4}
In this work, we propose a model-independent method for simultaneous constraining on Hubble constant and  post-Newtonian parameter using time-delay strong lensing systems and observational SNe Ia. To match the lensing and source redshifts of the four strong lensing systems analyzed by H0LiCOW program, we use GP regression to reconstruct distance from observations of SNe Ia instead of assuming a specific model. Although the observations of SNe Ia provide the unanchored or relative distance, strong lensing systems encode absolute distance. Thus, such dataset combinations can anchor cosmological distances.

Firstly, we directly use four posteriors of $D_{\Delta t}$ (inferred from lensing mass) and $D_{d}$ (inferred from dynamic mass) published by H0LiCOW program, combining with reconstructed unanchored SN dataset. For combination of all lenses, we find that the constraint result on PPN parameter is $\gp=0.89^{+0.17}_{-0.15}$ which demonstrates that GR is supported within observational uncertainties. The result on Hubble constant is $H_0=72.9^{+2.0}_{-2.3} \Mpc$, which is consistent with the time delay and time delay plus SN results \citep{2019PhRvL.123w1101C,2020MNRAS.498.1420W,2019A&A...628L...7T} made within specific cosmological models or with polynomial fitting of the distance relation.

Secondly, in order to avoid the influence of $H_0$ on GR testing, we do not assume any value for $H_0$, and consider a ratio of distance ${D_{\Delta t}}/{D^{'}_{d}}$ method to test PPN parameter. This method can independently give a constraint on $\gp$, because the distance ratio contains only velocity dispersion measurements and the dispersion measurements do not play a significant role in the $H_0$. The mean values of $\gp$ for four individual lenses have a little changes (though not significant) using ratio of distance method. However, the constraint result using the lens J1206 is $\gp=1.78^{+0.87}_{-0.67}$, which no longer supports that GR holds within observational uncertainty.

As a final remark, we point out that time-delay lenses plus observations of SN provide a quite promising and model independent method to test General Relativity. The uncertainty of model-independent analyses with such dataset combination can be comparable to the uncertainty of assuming specific models, while reducing possible biases. We also look forward to a large amount of future data, not only from strong lensing system, but also from the SN Ia, allowing us to further improve the measurements of $H_0$ and $\gp$.  In the future,
current surveys such as Dark Energy Survey (DES) \citep{2018MNRAS.481.1041T} and Hyper SuprimeCam Survey (HSC) \citep{2017MNRAS.465.2411M}, and the upcoming wide-area and deep surveys like the Large Synoptic Survey Telescope (LSST) \citep{2010MNRAS.405.2579O} and Euclid and WFIRST satellites \citep{2018PhR...778....1B,2018A&A...614A.103P} with much wider field of view and higher sensitivity will be able to discover and precisely localize a large number of lensed quasars and even other lensed sources, which will have well-measured time delays. The High-resolution imaging from space telescopes such as HST or ground-based adaptive optics will help to better model the stellar kinematic in lensing galaxies. In addition, SN Ia data will continue to improve as well,  playing an important role as a dense sampler of cosmic expansion history over a wide range of redshifts.
All these help us to carry out more accurate analysis of General Relativity, Hubble constant, and cosmology in our subsequent work.

\section{Acknowledgments}

This work was supported by National Natural Science Foundation of China under Grant Nos. 12203009, 12222302,  11973034. Liu. T.-H was supported by Chutian Scholars Program in Hubei Province (X2023007).  Liao. K was supported by Funds for the Central Universities (Wuhan University 1302/600460081).

\section{Data availability statements}
The data underlying this article will be shared on reasonable
request to the corresponding author.

\end{document}